\documentstyle[12pt]{article}
\newcommand{\be}{\begin{equation}}
\newcommand{\ee}{\end{equation}}
\newcommand{\ba}{\begin{eqnarray}}
\newcommand{\ea}{\end{eqnarray}}
\newcommand{\bb}{\begin{thebibliography}}
\newcommand{\eb}{\end{thebibliography}}
\newcommand{\ci}[1]{\cite{#1}}
\newcommand{\bi}[1]{\bibitem{#1}}
\newcommand{\lab}[1]{\label{#1}}

\begin{document}
%\phantom{.}
%\vspace{-2cm}

\begin{center}

{\large{Double spin asymmetry  in diffractive $Q \bar Q$ production at SMC
energies and quark-pomeron vertex structure.}}
\\
S.V.Goloskokov, \\
BLTP, JINR, Dubna, Russia
\footnote{Email: goloskkv@thsun1.jinr.dubna.su}\\

\vspace{.7cm}
\begin{abstract}
It is shown that the longitudinal
double spin asymmetry $A_{ll}$ in  polarized diffractive
$l p \to l p+Q \bar Q$ reaction depends strongly on the spin structure
of quark-pomeron vertex. The SMC spin facilities may be one of the best place
to
perform such experiment.
\end{abstract}

\end{center}

At present the study of diffractive processes has attracted considerable
interest due to the observation of high $p_t$ jets in diffractive
collisions \ci{ua8,h1}. These events can be interpreted as the
observation of partonic structure of pomeron \ci{ing}. However this
structure can be determined by quarks and gluons.
 It have been shown in papers \ci{lanj,coll} that the observes effects can
be predominated by quark structure function of pomeron which was confirmed
experimentally \ci{h1}.

Different models are used in investigations of pomeron exchange contribution.
First is the BFKL pomeron \ci{bfkl} which is based on the summation of
leading perturbative logarithms, second is the nonperturbative model of
two-gluon exchange \ci{low,la-na}.
Such "bare" pomeron exchange leads to the mainly imaginary helicity
conserving scattering amplitude because quark-pomeron coupling in this case
is like a C=+1 isoscalar photon (see e.g. \ci{la-na}). So, the analysis of
 pomeron structure functions using "bare" pomeron is equivalent in some
sense to the spin-average quark and gluon distributions inside pomeron.

However the spin structure of quark-pomeron vertex may be not so simple.
Factorization of the $qq$ amplitude was shown
into the spin-dependent large-distance part and the high-energy spinless
pomeron \ci{gol-pl}.
The quark-pomeron vertex  in the semi-hard region
$s \to \infty,\; |t|>1GeV^2$ \ci{gol-pl} where the perturbative theory can be
used has a form
\be
V_{qqP}^{\mu}(k,q)=\gamma_{\mu} u_0(q)+2 m k_{\mu} u_1(q) +2 k_{\mu}
/ \hspace{-2.3mm} k u_2(q) + i u_3(q) \epsilon^{\mu\alpha\beta\rho}
k_\alpha q_\beta \gamma_\rho \gamma_5+im u_4(q)
\sigma^{\mu\alpha} q_\alpha.    \lab{ver}
\ee
In (\ref{ver}) $u_i(q)$ are the vertex functions.
 Note that the structure of the quark-pomeron vertex function (\ref{ver}) is
drastically different from the  standard spinless pomeron.
Really, only the term
proportional to $\gamma_{\mu}$  corresponds to the standard helicity
conserving
pomeron and reflects the well-known fact that the spinless quark-pomeron
coupling is like a $C=+1$ isoscalar photon. The terms
$u_1(q)-u_4(q)$ lead to the spin-flip in the quark-pomeron vertex in contrast
to the term proportional to $u_0(q)$.  The functions
$u_1(q) \div u_4(q)$ at large $Q^2$ were calculated in perturbative QCD
\ci{gol4}. Their magnitudes are not very small.

As a result
the spin-dependent quark-pomeron vertex should modify different spin
asymmetries and lead to new effects in high energy diffractive reactions
which can be measured in future spin experiments in the RHIC at Brookhaven
\ci{bunce} for example.

The obtained structure of quark-pomeron vertex can be detected in $Q \bar Q$
diffractive production by the polarized protons. To show this we have been
estimated the longitudinal
double spin asymmetries in this reaction  as an example.  The
resulting asymmetry can reach $10 \div 12 \% $ \ci{golasy}.

However in $pp$ polarized diffractive reactions the obtained asymmetry
depends on practically unknown $\Delta g$ gluon structure function of proton.
In order to obtain more explicit results it is better to use $l p$ polarized
beams. We hope that the SMC spin facilities is one of the best place to
perform such experiment.

In what follows we shall analyse the longitudinal
double spin asymmetry  for polarized $l p \to l p+Q \bar Q$ reaction at SMC
energies which can be detected as $l p \to l p+ 2 Jets$
events. The standard set of kinematical variables looks as follows \ci{h1}
(see fig.1):
\ba
s=(p_l+p)^2,\; \;Q^2=-q^2,\;t=(p-p')^2 \nonumber
\\  y=\frac{pq}{p_lq},\;x=\frac{Q^2}{2pq},\;
\beta=\frac{Q^2}{2q(p-p')},\;x_p=\frac{q(p-p')}{qp},
\ea
where $p_l,p'_l$ and $p, p'$ are the initial and final lepton and proton
momenta respectively, $q=p_l-p'_l$.

In calculations we shall use the hypothesis that pomeron
couples to single quark (or antiquark) for simplicity. The integration over
all $ Q \bar Q $ phase space will be performed.
We shall omit in calculations the non-log corrections like $\beta$ that is true
in the case when the energy of $Q\bar Q$ system is large with respect to the
momenta transfer $t$ and $Q^2$. We shall omit non-log $m_Q^2/|t|$ corrections
too. The last simplification modifies numerical results within $5 \div 10$ \%
 accuracy.

The resulting longitudinal
double spin asymmetry defined via

\be
A_{ll}=
\frac{\Delta \sigma}{\sigma}=\frac{\sigma(^{\rightarrow} _{\Rightarrow})
+\sigma(^{\leftarrow}_{\Leftarrow})
-\sigma(^{\leftarrow} _{\Rightarrow})- \sigma(^{\rightarrow}_{\Leftarrow})}
{\sigma(^{\rightarrow} _{\Rightarrow})
+ \sigma(^{\leftarrow}_{\Leftarrow})
+\sigma(^{\leftarrow} _{\Rightarrow})+\sigma(^{\rightarrow}_{\Leftarrow})}.
\lab{asydef}
\ee
The main contributions to $A_{ll}$ asymmetry in discussed region
are determined by $u_0$ and $u_3$ structures in (\ref{ver}).
It can be written in the form:
\be
A_{ll}=-x_P y\frac{A_{00} u_0^2+A_{03} |t| u_0 u_3}
{S_{00} u_0^2+S_{03} |t| u_0 u_3+S_{33} |t|^2 u_3^2}. \lab{asy}
\ee
where
\ba
A_{00}=2 Q^2 (2-y) (\ln(\frac{|t|}{m_Q^2})-3), \nonumber \\
A_{03}=- Q^2 (2-y) ( 2 \ln( \frac{|t|}{m_Q^2})-3),     \lab{a}
\ea
and
\ba
S_{00}= 2 [ 2 |t| (1-y) \ln( \frac{Q^2}{|t|\beta}) +
 Q^2 y (2-y) \ln(\frac{|t|}{m_Q^2}) ],\hspace{3cm}\nonumber\\
S_{03}=- 2 [ 2 |t| (1-y)( \ln( \frac{Q^2}{|t|\beta})-1) +
 Q^2 y (2-y) (\ln(\frac{|t|}{m_Q^2})+1) ], \hspace{2cm}\nonumber \\
S_{33}= [ Q^2 y (2-y)\ln( \frac{Q^2}{|t|\beta})+
4 |t| (1-y)( \ln( \frac{Q^2}{|t|\beta})-1) +
 Q^2 y (2-y) (\ln(\frac{|t|}{m_Q^2})+2) ].
\ea
 Note that $u_3<0$.

It is easy to see from(\ref{a}) that $\Delta \sigma$ has different signs for
light ($m_Q \sim .005 GeV$) and heavy (for charm  $m_Q \sim 1.3 GeV$) quarks
production. Moreover $\Delta \sigma$ is proportional to $Q^2$ and $\sigma$
has a more complicated structure. As a result the asymmetry must increase
with $Q^2$. The obtained asymmetry is equal to zero for $x_p=0$ and it is
better to analyse it at $x_p=0.1 \div 0.2$.

Our predictions for $A_{ll}$ asymmetry for energy $\sqrt s=20GeV$  estimated on
the basis of perturbative results for vertex functions for $y=0.5$ and
$x_p=0.2$ for standard quark pomeron vertex ($u_0$z terms in (\ref{asy}) only)
and spin-dependent quark pomeron are shown in fig. 2,3.
On fig.2 the $|t|$ dependence of $A_{ll}$ for fixed $Q^2=10GeV^2$
($\beta=0.25$) is shown.
On fig.3 one can see the $Q^2$ dependence of $A_ll$ for fixed $|t|=3GeV^2$.
The obtained asymmetry is not small and depend strongly on the spin structure
of quark-pomeron vertex.
For spin-dependent quark-pomeron vertex $A_ll$ asymmetry is smaller by
factor 2 because the $\sigma$ in (\ref{asydef}) is larger in this case.
Asymmetry decrease with $|t|$ growth and increase with growth of $Q^2$.
It is positive
for $C$ quark production and negative  for light quark production.

 The estimations shows that total integrated cross section of light quark
production in $lp$ reaction is about $0.2 \div 1$ nb \ci{lanj,rysk}.
Our calculations shows that the cross section for $C$ quark production must
be smaller by factor $3 \div 10$.

The discussed here spin-dependent contributions to the quark-pomeron
and hadron-pomeron vertex functions modify different spin
asymmetries and lead to new effects in high energy diffractive reactions
which can be measured in spin experiments at future accelerators.

To summarize, we have present in this letter the perturbative QCD analyses of
longitudinal double spin asymmetry in diffractive 2-jet production in $lp$
processes. The model prediction shows that the $A_{LL}$ asymmetry can be
measured and the information about the spin structure of the quark-pomeron
vertex can be extracted.
It should be emphasized that the obtained here spin effects are
completely determined at fixed momenta transfer by the large-distance
contributions in quark (gluon) loops. So, they have a nonperturbative
character. The investigation of spin effects in diffractive reactions is an
important test of spin sector of QCD at large distance.

The author express his deep gratitude to A.V.Efremov, V.G.Krivokhijine,
W-D.Nowak, S.B.Nurushev, I.A.Savin, O.V.Terjaev for fruitful discussion.
%\newpage

\newpage
{\bf{Figure captions}}\\ {\bf{}}
{\bf{Fig.1}} The diffractive $Q \bar Q$ production in $lp$ reaction.\\
{\bf{Fig.2}} The $|t|$ dependence of $A_{ll}$ asymmetry of light and heavy (C)
quarks production at fixed $Q^2=10GeV^2$. Solid line -for standard;
dot-dashed line -for spin-dependent quark-pomeron vertex.\\
{\bf{Fig.3}} The $Q^2$ dependence of $A_{ll}$ asymmetry of light and heavy (C)
quarks production at fixed $|t|=3GeV^2$. Solid line -for standard;
dot-dashed line -for spin-dependent quark-pomeron vertex.

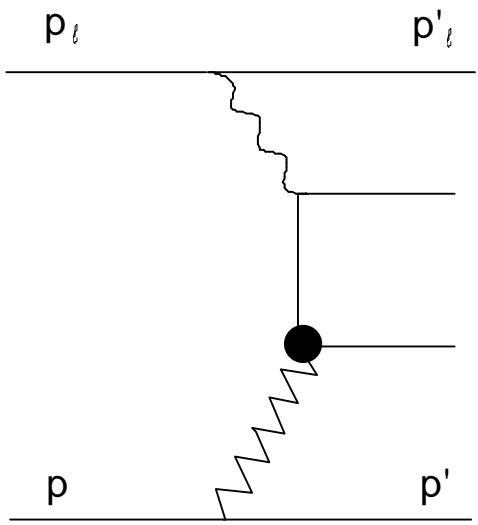
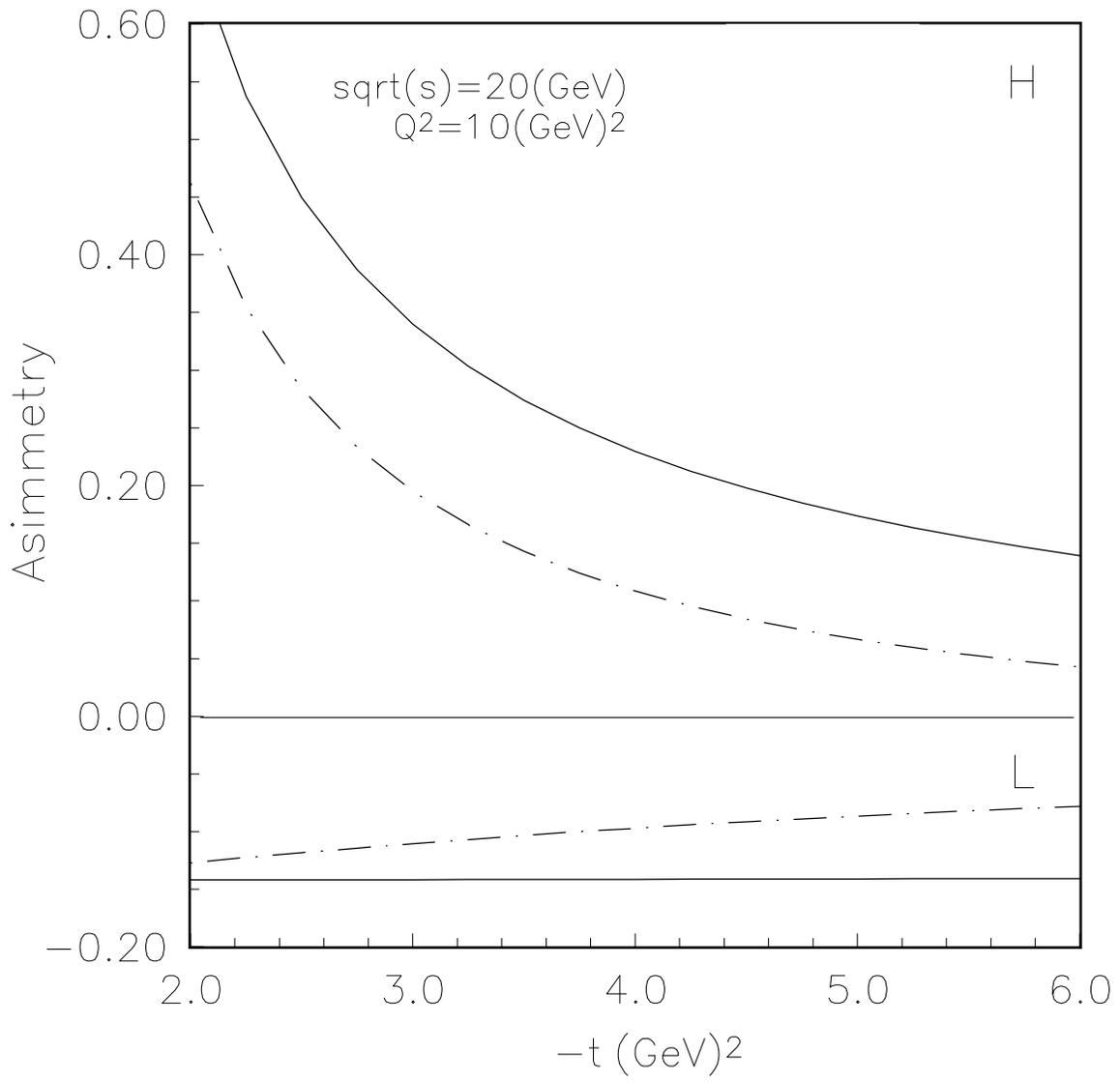
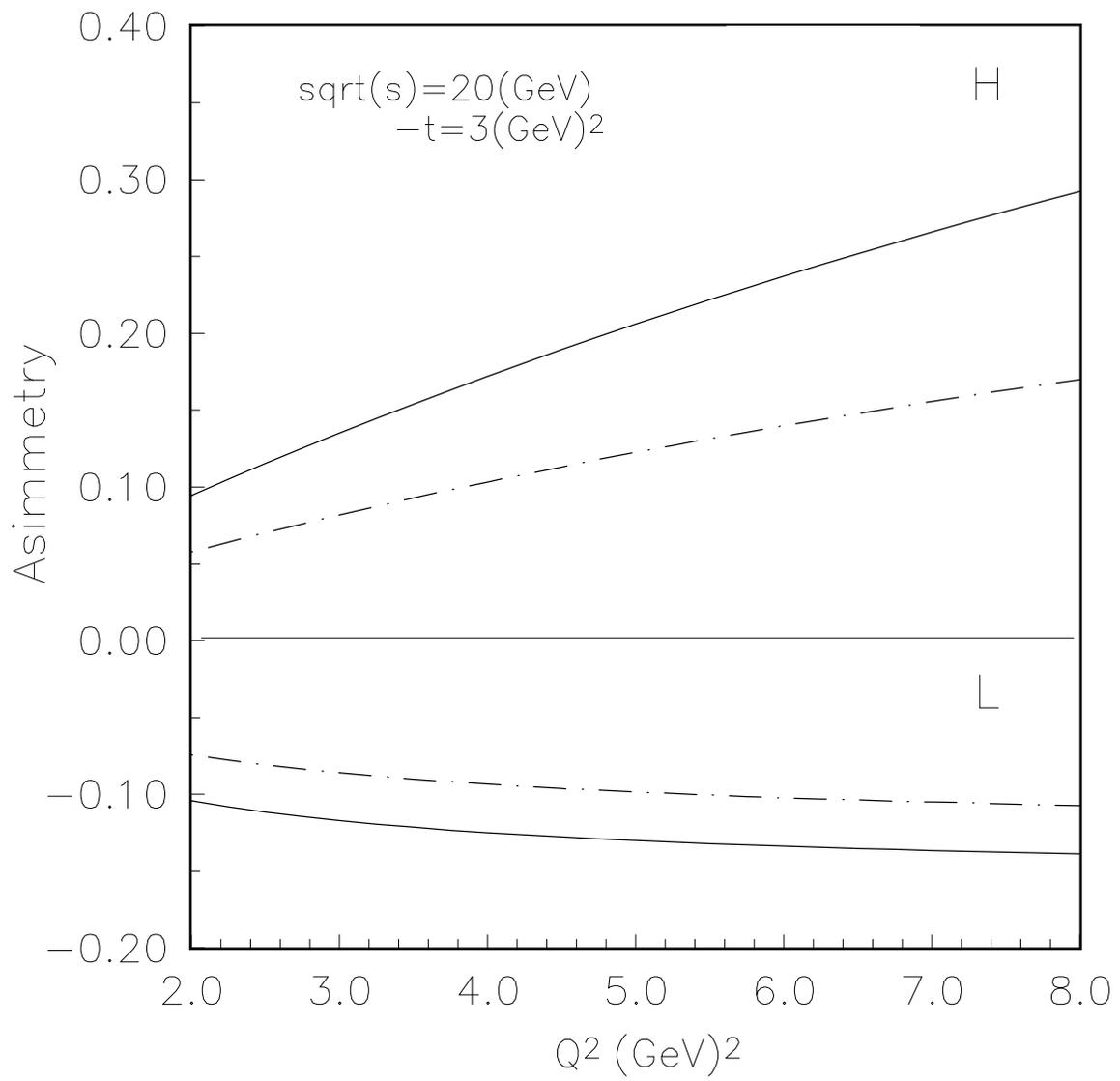
\end{document}